\documentclass[12pt]{article}
\pdfoutput=1
\usepackage{setspace}
\usepackage{amssymb}
\usepackage{amsmath}
\usepackage{bm}
\usepackage{times}
\usepackage[]{graphicx}
\usepackage{scicite}
\usepackage[labelfont=bf]{caption}
\usepackage{times}

\newcounter{lastnote}
\newenvironment{scilastnote}{%
\setcounter{lastnote}{\value{enumiv}}%
\addtocounter{lastnote}{+1}%
\begin{list}%
{\arabic{lastnote}.}
{\setlength{\leftmargin}{.22in}}
{\setlength{\labelsep}{.5em}}}
{\end{list}}

\title{{\bf Flagellar Synchronization Through \\ Direct Hydrodynamic Interactions}}

\author
{Douglas R. Brumley,$^{1,2\dag}$ Kirsty Y. Wan,$^{1}$, Marco Polin,$^{1,3\dag}$ Raymond E. Goldstein$^{1\ast}$\\
\\
\normalsize{$^{1}$Department of Applied Mathematics and Theoretical Physics, University of Cambridge,}\\
\normalsize{Wilberforce Road, Cambridge CB3 0WA, United Kingdom}\\
\normalsize{$^{2}$Department of Civil and Environmental Engineering, Massachusetts Institute of Technology,}\\
\normalsize{77 Massachusetts Avenue, Cambridge, Massachusetts 02139, USA}\\
\normalsize{$^{3}$Physics Department, University of Warwick,}\\
\normalsize{Gibbet Hill Road, Coventry CV4 7AL, United Kingdom}\\
\\
\normalsize{$^\ast$To whom correspondence should be addressed; E-mail: R.E.Goldstein@damtp.cam.ac.uk}\\
\normalsize{$^\dag$Present address}
}

\date{}

\begin{document} 

\baselineskip18pt


\maketitle

\vspace{-12pt}

%\begin{sciabstract}

\baselineskip16pt
{\bf Microscale fluid flows generated by ensembles of beating eukaryotic flagella are crucial to fundamental processes such as development, 
motility and sensing. Despite significant experimental and theoretical progress, the underlying physical mechanisms behind this striking 
coordination remain unclear. Here, we present a novel series of experiments in which the flagellar dynamics of two micropipette-held somatic cells of 
\textit{Volvox carteri}, with measurably different intrinsic beating frequencies, are studied by high-speed imaging as a function of
their mutual separation and orientation. From analysis of beating time series we find that the interflagellar coupling, 
which is constrained by the lack of chemical and mechanical connections between the cells to be purely hydrodynamical, exhibits a
spatial dependence that is consistent with theoretical predictions.  At close spacings it 
produces robust synchrony which can prevail for thousands of flagellar beats, while at increasing separations this synchrony is 
systematically degraded by stochastic processes. Manipulation of the relative flagellar orientation reveals the existence 
of both in-phase and antiphase synchronized states, which is consistent with dynamical theories. Through dynamic flagellar tracking with exquisite spatio-temporal precision, we quantify changes in beating waveforms that result from altered coupling configuration and distance of separation. The experimental results of this study prove unequivocally that flagella coupled solely through a fluid medium can achieve robust synchrony despite significant differences in their intrinsic properties.}
%\end{sciabstract}

\baselineskip18pt

% In setting up this template for *Science* papers, we've used both
% the \section* command and the \paragraph* command for topical
% divisions.  Which you use will of course depend on the type of paper
% you're writing.  Review Articles tend to have displayed headings, for
% which \section* is more appropriate; Research Articles, when they have
% formal topical divisions at all, tend to signal them with bold text
% that runs into the paragraph, for which \paragraph* is the right
% choice.  Either way, use the asterisk (*) modifier, as shown, to

\section*{Introduction}

Despite the elegance and apparent simplicity of the eukaryotic flagellum and its shorter ciliary version, the collective motions 
exhibited by groups of these organelles and the resultant fluid flows are far from trivial. For example, the unicellular biflagellate 
alga \textit{Chlamydomonas} executes diffusive `run-and-turn' locomotion\cite{Polin:2009kx,Goldstein:2009vn} through 
stochastic switching between synchronized and unsynchronized swimming gaits -- a process which could enhance searching efficiency
%\cite{Benichou2005} 
and assist in the avoidance of predators\cite{Stocker:2009kx}. Ensembles of cilia and flagella 
exhibit stunning temporal coordination, generating flows that transport mucus and expel pathogens
 %such as \textit{Pseudomonas aeruginosa} from the human respiratory tract
 \cite{Button:2012}, establish the left-right asymmetry in developing mammalian embryos%\cite{Cartwright:2004,Nonaka2002}
\cite{Nonaka2002}, and transport ova in human fallopian tubes\cite{Lyons:2006}. \\
\indent The origin of flagellar synchronization has been the subject of intense theoretical investigation for many decades. In 
1951, Taylor\cite{Taylor:1951} showed that viscous dissipation of nearby swimming sheets is minimised in the in-phase 
state, and since then a myriad of increasingly complex models of flagellar synchronization have been proposed. Hydrodynamically coupled 
filaments or chains with various internal driving forces exhibit a general tendency towards 
synchrony\cite{Machin:1963,Gueron:1997,Guirao:2007fk,Elgeti:2013,Yang:2008}. At the same time, 
minimal models of coupled oscillators in viscous fluids\cite{Brumley:2012,Vilfan:2006uq,Niedermayer:2008fk,Uchida:2011kx,Uchida:2012fk} 
offer great insight into the emergence of metachronal coordination.
Such models have been investigated experimentally with light driven microrotors\cite{DiLeonardo12}, rotating paddles\cite{Qian:2009} and colloids in optical tweezers\cite{Kotar:2010}, and have also given rise to interpretations of the synchrony and coupling interactions between pairs of flagella of the model alga \textit{Chlamydomonas} \cite{Goldstein:2009vn}.
Although experimentally-derived coupling strengths between \textit{Chlamydomonas} flagella are consistent with hydrodynamic predictions \cite{Goldstein:2011_emergence}, it is difficult to establish with certainty the origins of the coupling mechanism due to the likely presence of biochemical and elastic couplings of as yet unquantified strength between flagella.
In order to disentangle the hydrodynamic from the intracellular contributions to flagellar synchronization we conducted a series of experiments in which two physically separated flagellated cells, which exhibit distinct intrinsic beating frequencies in isolation, are coupled solely and directly through the surrounding fluid.  
Owing to the natural distribution of beating frequencies of the flagella of its surface somatic cells, the colonial 
alga \textit{Volvox carteri} is ideally suited to this purpose. 
Each somatic cell possesses two flagella which beat in perfect synchrony, facilitating their treatment as a single entity, henceforth referred to as the {\it flagellum}.
Somatic cells were isolated from adult \textit{Volvox} colonies and held with micropipettes at a controllable separation $d$ (Fig. 1A). 
The spatial and orientational degrees of freedom associated with this configuration enabled comprehensive analysis over a wide range of hydrodynamic coupling strengths. We found that closely-separated pairs of cells can exhibit robust phase-locking for thousands of beats at a time, despite a discrepancy in their intrinsic frequencies of as much as 10\%. Both in-phase and antiphase configurations were observed, depending on the alignment of the directions of flagellar propulsion. Furthermore, with increasing interflagellar spacing we observed for each flagellum a marked change in the beating waveform, a key finding that lends support to models of synchronization that rely on waveform compliance to achieve phase-locking.

\section*{Results}

\textit{One cell.}---We begin by characterising the flow generated by a single beating flagellum. 
Time-dependent as well as time-averaged flow fields were obtained using particle image velocimetry (Fig.~2).
A point force $\bm{F}$ (Stokeslet) exerted on a viscous fluid at a point $\bm{x}_0$ produces a velocity 
field \cite{BlakeAndChwang:1973} of the form $u_i = \frac{F_j}{8 \pi \mu} \big( \delta_{ij}/r + r_i r_j/r^3 \big)$, 
where the vector $\bm{r} = \bm{x} - \bm{x}_0$, $r = |\bm{r}|$ and $\delta_{ij}$ is the Kronecker delta. 
The time-dependent flow field shown in Fig.~2A is well described by that of a Stokeslet with time-varying position, direction and magnitude (Fig.~S6, supplementary materials). 
Examining the time-averaged velocity field (Fig.~2B, obtained by averaging data from four cells), we see that for distances larger than 20~$\mu$m from the flagellar tip, both \textit{upstream} (red) and \textit{downstream} (blue) components of the flow obey a Stokeslet decay ($u\sim 1/r$)  (Fig.~2C). 
This trend is maintained over a range consistent with the distances sampled for our two-cell experiments (below).

\textit{Two cells.}---To investigate the effect of hydrodynamic coupling on \textit{pairs} of flagella, we captured pairs of cells and aligned them so that their flagellar beating planes coincided (Fig.~1A). 
At each cell-cell separation $d$, we recorded flagellar dynamics over $104$~s, and extracted flagellar phases $\phi_{1,2}$ from Poincar\'e sectioning of the dynamics\cite{Polin:2009kx,Goldstein:2009vn} by monitoring the signal in respective interrogation regions (Fig.~1B). 
The measured interflagellar phase difference, defined by $\Delta = (\phi_1 - \phi_2)/2\pi$, is shown in Fig.~1C for one pair of cells at four different spacings. 
We measured beat frequencies $\omega_1$ and $\omega_2$ for the two flagella, and define $\delta \omega = \omega_1 - \omega_2$ to be the intrinsic frequency difference.
Calling $L = d/l$ the cell-cell separation normalised by the flagellar length $l$, Fig.~1C shows that 
for $L\gg 1$ hydrodynamic coupling is negligible and 
$\Delta(t)$ drifts approximately linearly with time depending on $\delta \omega$ (8.2\% here). For intermediate values of 
$L$, the flagella exhibit short periods of synchrony interrupted by brief phase slips. 
However, when the same cells are brought closer to each other, they phase-lock for the entire duration of the experiment.
This conclusively demonstrates that robust and extended flagellar synchronization can arise in physically separated cells purely through the action of hydrodynamics. For different pairs of cells, a similar behaviour is observed. 

Next we examine in detail the experimental time series $\Delta(t)$. 
%Consider first the synchronous periods within the full time series of Fig.~1C. 
Fluctuations about the phase-locked states $\Delta_0$ (Fig.~1D) are Gaussian with a variance proportional to $L$, as seen by rescaling as $(\Delta-\Delta_0)/L^{1/2}$ (Fig.~1E).
Gaussian fluctuations suggest a description of the dynamics of $\Delta(t)$ based on a Langevin equation with an effective potential $V(\Delta)$ having a quadratic minimum at $\Delta_0$. We then write 
\begin{equation}
\dot{\Delta} = -V'(\Delta) + \xi(t), \label{model_equation}
\end{equation}
%Fluctuations about the phase-locked states (Fig.~1D) are Gaussian with a variance  proportional to $L$, as evidenced by the data collapse in the variable $(\Delta-\Delta_0)/L^{1/2}$ (Fig.~1E).  
%This Gaussian behaviour implies that a Langevin equation for the phase difference is characterised by an effective potential $V(\Delta)$ that has a quadratic 
%minimum near $\Delta_0$, hence of the form
%\begin{equation}
%\dot{\Delta} = -V'(\Delta) + \xi(t), \label{model_equation}
%\end{equation}
where $V(\Delta) = -\delta \nu \Delta + U(\Delta)$. The value $\delta \nu$ is the intrinsic frequency difference for the two phase oscillators, 
$U$ an effective potential which has period one in $\Delta$, and $\xi(t)$ is a Gaussian white noise term satisfying $\langle \xi(t) \rangle=0$ 
and $\langle \xi(t)\xi(t') \rangle = 2T_{\text{eff}} \delta(t-t')$. To leading order $U = -\epsilon \cos (2 \pi \Delta)$, where 
$\epsilon$ is the interflagellar coupling strength.  
The observed dependence on $L$ of the distribution of $\Delta$ fluctuations is a natural consequence of Eq.~\eqref{model_equation} if $\epsilon \propto 1/L$. We test this scaling below. 
Intraflagellar biochemical noise leads to stochastic transitions between adjacent minima of the tilted 
washboard potential $V(\Delta)$ \cite{Polin:2009kx,Goldstein:2009vn}.  
For each  movie, the autocorrelation of $\Delta$ is used to extract the model parameters $(\epsilon, \delta \nu, T_{\text{eff}})$ as described previously \cite{Polin:2009kx,Goldstein:2009vn}.

Cells aligned so that their power strokes point in the same direction (as in many ciliates) exhibit 
\textit{in-phase} (IP) synchrony ($\Delta \simeq 0$), indicating a coupling strength $\epsilon > 0$. Rotation of pipette P$_1$ (Fig.~1B) 
by $180^{\circ}$ so that the power strokes are opposed (as in the {\it Chlamydomonas} breaststroke) changes the sign of the 
coupling strength and gives rise to \textit{antiphase} (AP) synchronization ($\Delta \simeq 1/2$), in agreement with
theory \cite{Leptos:2013}. 
%In many cases the phase-locked value of $\Delta$ deviates slightly from these (half) integer values -- a lag arising from the difference in intrinsic flagellar frequencies. 
Figure~3A depicts the nondimensionalised coupling strength $\kappa=\epsilon/\bar{\omega}$ for all experiments, 
where  $\bar{\omega}$ is the average beat frequency across all experiments for a given pair of cells. The dependence on the 
interflagellar spacing $|\kappa | \propto L^{-1}$ is consistent with the intrinsic flagellar flow field presented in Fig.~2. For both the 
in-phase and antiphase configurations, we fit $|\kappa| = k\times L^{-1}$ finding $k_{\rm IP}=0.016$ and $k_{\rm AP}=0.014$ respectively. 
At a given $L$, IP pairs exhibit on average a marginally stronger coupling than AP ones, possibly due  to the fact that 
flagella in IP are on average closer together than in AP.
%The in-phase configuration exhibits a slightly higher coupling strength than the antiphase mode, 
%likely  due to the fact that flagellar filaments in the former state are on average closer together for a given value of $L$.  
%
The average values of the other model parameters are $\langle T_{\text{eff}} / \bar{\omega}\rangle = 0.005 \pm 0.003$ and 
$\langle \delta \nu / \bar{\omega} \rangle = 0.058 \pm 0.033$, with $\langle \bar{\omega} \rangle = 33.0$~Hz. 

The average \textit{measured} flagellar frequency $\omega$ for the two cells in each experiment is shown in 
Fig.~3B, nondimensionalised by the average value for each cell $\bar{\omega}_{\text{cell}}$ across different videos. Figure~3C illustrates the measured frequency difference as 
a function of $L$. The data exhibit a bifurcation near $L = 1$, beyond which phase drifting occurs over time. Integration 
of Eq.~\eqref{model_equation} in the absence of noise yields a predicted value for the observed frequency difference in 
terms of the model parameters, $\delta \omega / \delta \omega_{\text{far}} = \sqrt{1-(2\pi \epsilon / \delta \nu)^2}$. 
The orange curve in Fig.~3C illustrates this prediction, calculated using the average extracted model parameters. The average phase drift in the presence of noise is also shown in green\cite{Risken}. It is evident that noise plays an important role in determining the observed location of the bifurcation point.

\textit{Waveform characteristics.}---Although coupling is established purely through hydrodynamic interactions, the process of synchronization hinges on the ability of the flagella to respond differentially to varying external flows. For sufficiently strong coupling, nonidentical cells can adopt a common phase-locked frequency through perturbing one another from their intrinsic limit cycles. Indeed, models of coupled flagella involving hydrodynamically coupled semiflexible filaments\cite{Gueron:1997,Guirao:2007fk,Elgeti:2013,Yang:2008} show a tendency towards metachronal coordination, though the precise role that flexibility plays in facilitating synchrony is unknown. Minimal models in which spheres are driven along flexible trajectories\cite{Niedermayer:2008fk,Brumley:2012} reveal that deformation-induced changes in the phase speed can facilitate synchrony. However, functional variations in the intrinsic flagellar driving forces could lead to synchrony even for fixed beating patterns\cite{Uchida:2011kx,Uchida:2012fk}. 

Through dynamic tracking \cite{wan14}, we followed the evolution of the flagellar waveforms for several thousand consecutive beats. One example is shown in Fig.~4A where the extracted waveform is shown at various stages through the beating cycle, along with the logarithmically-scaled residence time plots. The same pair of flagella is compared at close and far cell-cell separations (7.3~$\mu$m and 72.2~$\mu$m respectively).
In order to quantity flagellar waveform changes as the cells are brought closer together, we define three angles $x_a$, $x_b$, $x_c$ (radians) with respect to the cell body axis (Fig.~4B). Figure~4C shows the temporal evolution of these angles for the right flagellum, corresponding to the close (red), intermediate (green) and wide (blue) separations. In particular, the most significant difference is observed in the $x_c$ component (distal part of the flagellum). Similar results are found for the other cell, indicating that the interaction is mutual. Figure~4 demonstrates that accompanying the robust hydrodynamic phase-locking is a change in the flagellar waveform. For the first time, we have shown by systematically varying the cell-cell spacing that each cell can directly alter the beating profile of its neighbour simply through hydrodynamic interactions. Extracted time-dependent waveforms can also be used to estimate the instantaneous force exerted by the flagellum on the fluid, the results of which agree well with forces inferred from the measured flow fields (supplementary materials).

\section*{Discussion}

Understanding the mechanisms giving rise to robust phase-locking of flagella can be broken down into two distinct components, namely (i) identification of physical or chemical coupling between the flagella and (ii) characterisation of the response of each flagellum subject to these external stimuli. Theoretical studies show that cell body rocking of freely swimming \textit{Chlamydomonas} can induce synchrony\cite{Geyer2013,Friedrich:2012fk}, and experimental investigations of such cells hint that hydrodynamic interactions between the flagella and cell body could be important for locomotion\cite{Kurtuldu2013}. At the same time, the synchronization properties of immobilised \textit{Chlamydomonas} cells are consistent with models in which the flagella interact purely hydrodynamically\cite{Polin:2009kx,Goldstein:2009vn,Goldstein:2011_emergence}. However, intracellular calcium fluctuations in \textit{Chlamydomonas} are known to affect the flagellar dynamics\cite{Yoshimura2003,Leptos:2013}, so these studies cannot exclude the possibility that flagellar synchrony is regulated primarily by chemical or other non-mechanical means. Indeed, Machemer showed that membrane voltage affects the metachronal wave direction in \textit{Paramecium}\cite{Machemer:1972ys}. 
In the present experiment, cells are held with micropipettes in order to eliminate rocking of the cell body. Through physical separation of the cells, we preclude all chemical coupling of the two flagella other than the possible advection of molecules by the flow. 
The fact that the coupling strength for IP and AP pairs is almost identical despite the pronounced difference between the associated flows rules out coupling via chemical means. In these experiments hydrodynamics alone is responsible for the interflagellar coupling.

Two spheres of radius $a$ in an unbounded fluid of viscosity $\mu$, driven along circular trajectories of variable radius (stiffness $\lambda$) are able to synchronize their motions through flow-induced changes in their respective phase speeds\cite{Niedermayer:2008fk}. Calculating the spring stiffness $\lambda = R/l^3$ in terms of the flagellar bending rigidity $R$ and length $l$ yields the scaled coupling strength for this dynamical system, $\kappa_{\text{spheres}} = (27 \mu \pi a^2 l^2  \bar{\omega} / 2R) \times L^{-1}$. Estimating\cite{Niedermayer:2008fk} $a=0.1~ \mu$m and $R=4\times10^{-22}~$Nm$^2$, and using measured values of the other parameters from the present experiment ($\langle \bar{\omega} \rangle = 33.0$~Hz and $\langle l \rangle=20.1~\mu$m), we obtain $\kappa_{\text{spheres}} = 0.014 \times L^{-1}$. This minimal model, in which synchronization is facilitated through hydrodynamic interactions, compares favourably with the measured flagellar coupling strengths presented in Fig.~3.

\section*{Conclusions}

The experimental study presented in this letter reveals unambiguously the importance of hydrodynamics in achieving flagellar synchronization. Physical separation of the cells precludes any form of chemical or direct mechanical coupling, leaving hydrodynamic interactions as the only mechanism through which synchronization can occur. The process of phase-locking is extremely robust, with cells sufficiently close to one another exhibiting perfect synchrony for thousands of consecutive beats. Accompanying this synchrony is a characteristic shift in the flagellar waveform. The extracted interflagellar coupling strength is consistent with hydrodynamic predictions and the measured flow fields of individual cells. Additional experiments were undertaken using the uniflagellar mutant of the unicellular alga \textit{Chlamydomonas}. Although its flagellum is shorter and its waveform is different to that of \textit{Volvox}, we also observed hydrodynamic phase-locking in these experiments. Owing to the ubiquity and uniformity in the structure and function of flagella in various eukaryotic species, the results of the present study are expected to be of significant value in a wide range of other systems and studies.

% Your references go at the end of the main text, and before the
% figures.  For this document we've used BibTeX, the .bib file
% scibib.bib, and the .bst file Science.bst.  The package scicite.sty
% was included to format the reference numbers according to *Science*
% style.

%\newpage 

%\begin{thebibliography}{}

% Following is a new environment, {scilastnote}, that's defined in the
% preamble and that allows authors to add a reference at the end of the
% list that's not signaled in the text; such references are used in
% *Science* for acknowledgments of funding, help, etc.

\begin{scilastnote}
\item {\bf Acknowledgements:} This work was supported by the EPSRC, ERC Advanced Investigator Grant $247333$ (REG), and 
a Senior Investigator Award from the Wellcome Trust (REG). An EPSRC Postdoctoral Fellowship (MP) contributed 
to the initial stages of the work and the Human Frontier Science Program (DRB) to the latter stages. 
\end{scilastnote}

% For your review copy (i.e., the file you initially send in for
% evaluation), you can use the {figure} environment and the
% \includegraphics command to stream your figures into the text, placing
% all figures at the end.  For the final, revised manuscript for
% acceptance and production, however, PostScript or other graphics
% should not be streamed into your compliled file.  Instead, set
% captions as simple paragraphs (with a \noindent tag), setting them
% off from the rest of the text with a \clearpage as shown  below, and
% submit figures as separate files according to the Art Department's
% instructions.

\clearpage

%\noindent {\bf Fig. 1.} Please do not use figure environments to set
%up your figures in the final (post-peer-review) draft, do not include graphics in your
%source code, and do not cite figures in the text using \LaTeX\
%\verb+\ref+ commands.  Instead, simply refer to the figure numbers in
%the text per {\it Science\/} style, and include the list of captions at
%the end of the document, coded as ordinary paragraphs as shown in the
%\texttt{scifile.tex} template file.  Your actual figure files should
%be submitted separately.
%

\newpage
\begin{figure*}
%\hspace{1.5cm} 
\begin{center}
\includegraphics[clip=true,width=0.8\columnwidth]{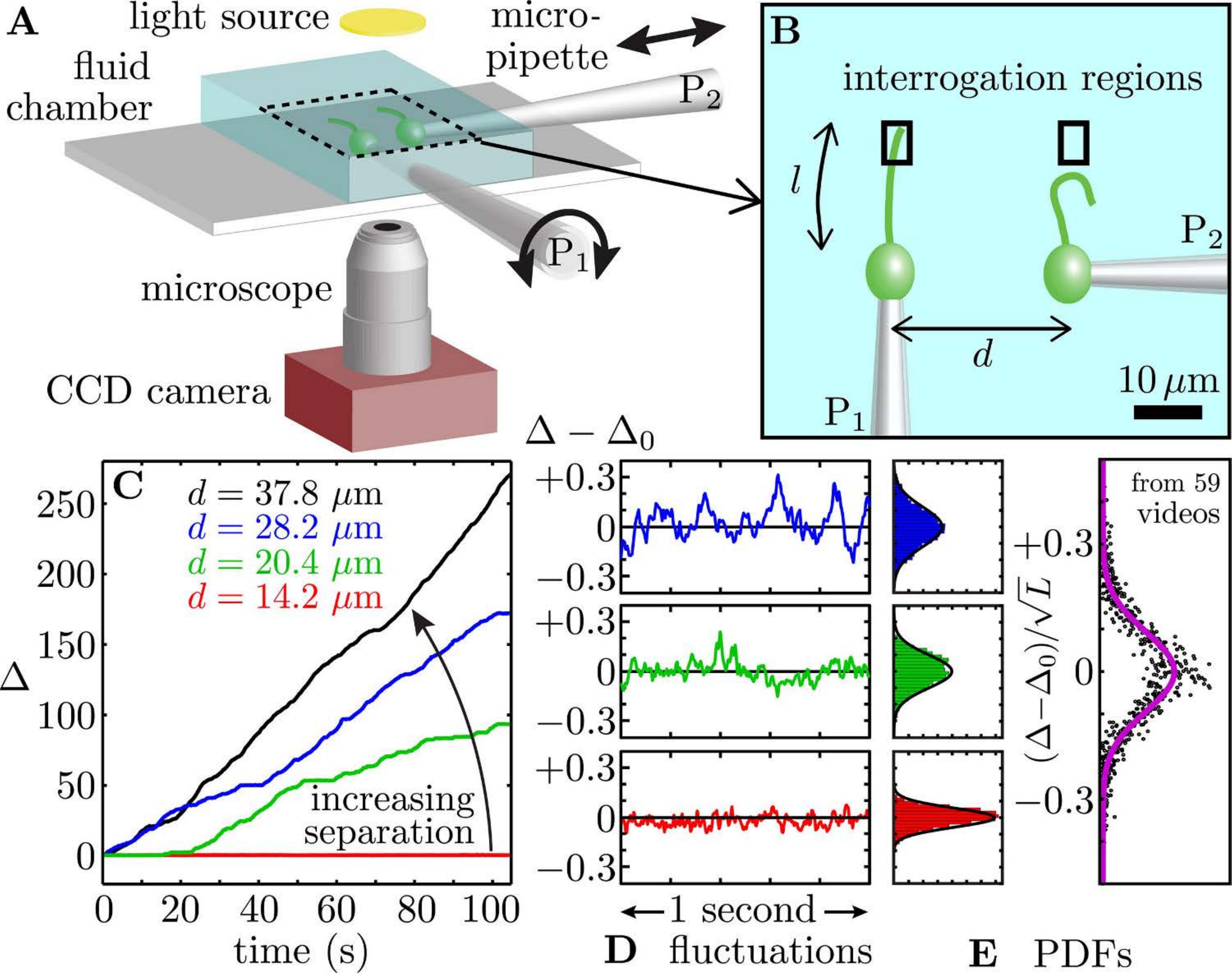}
\end{center}
\caption{\textbf{Synchronized pairs of beating flagella}. {\bf (A)} Experimental apparatus and {\bf (B)} cell configuration. 
{\bf (C)} Extracted phase difference $\Delta= (\phi_1 - \phi_2)/2\pi$ at four different interflagellar spacings, {\bf (D)} fluctuations during phase-locked periods around the average phase lag, $\Delta_0$, and {\bf (E)} the fluctuations' probability distribution functions (PDFs), each cast in terms of the rescaled separation-specific variable $(\Delta-\Delta_0)/\sqrt{L}$. Solid lines represent Gaussian fits.}
\end{figure*}

\newpage
\begin{figure*}
%\hspace{1cm}
\begin{center}
\includegraphics[clip=true,width=0.8\columnwidth]{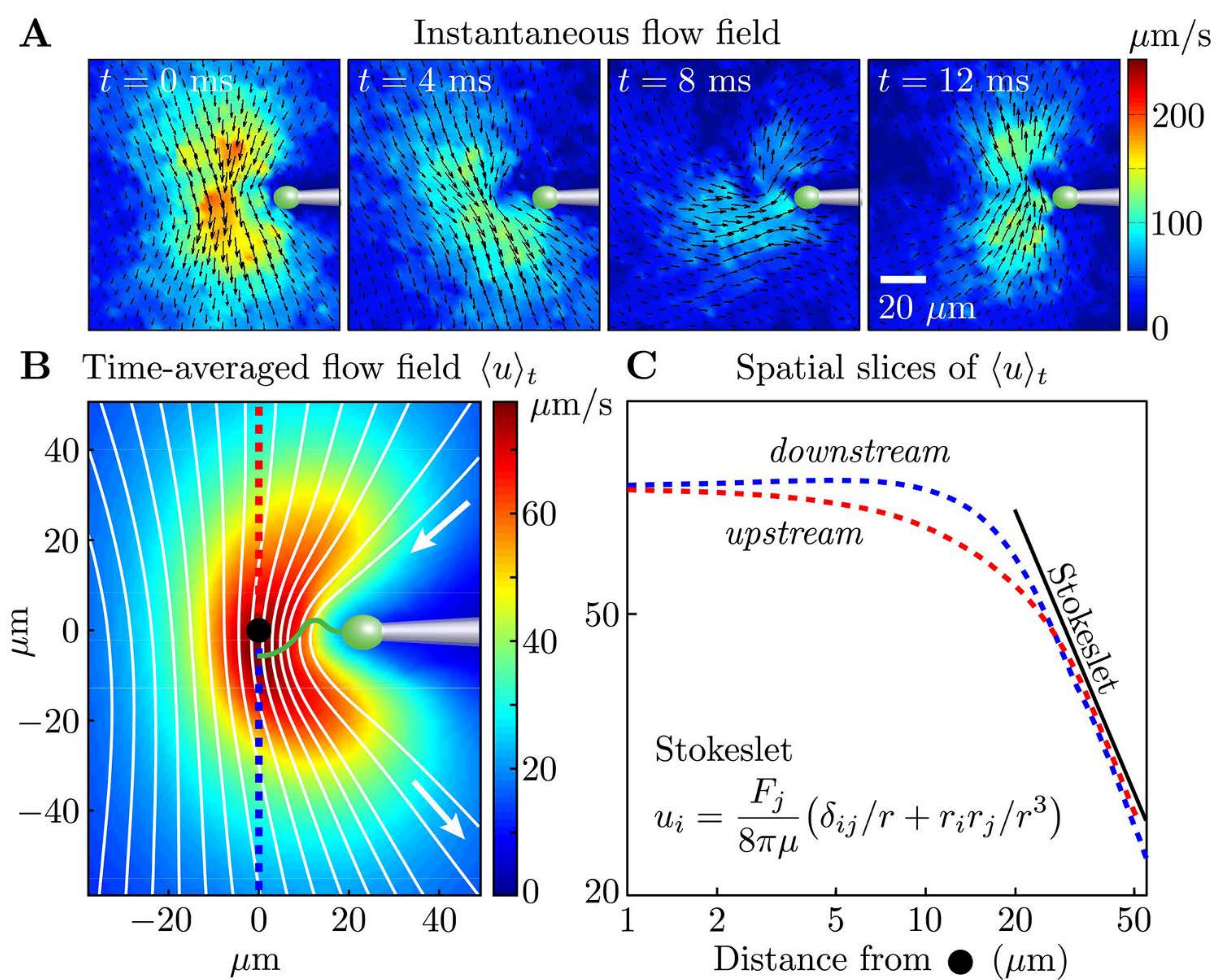}
\end{center}
\label{fig:fig2}
\caption{\textbf{Flagellar flow field.} {\bf (A)} Time-dependent flow field for an individual cell measured using particle image velocimetry. Results are shown for the first half of the beating cycle. {\bf (B)} Time-averaged flow field $\langle u \rangle_t = 1/\tau \int_0^{\tau} |\bm{u}(\bm{x},t)| dt$ (averaged across 4 cells with $\tau \sim 1000$ beats for each). The velocity magnitude (colour) and streamlines (white) are shown. {\bf (C)} Velocity magnitude \textit{upstream} (red) and \textit{downstream} (blue) of the origin (black dot in B).}
\end{figure*}

\newpage
\begin{figure*}
%\hspace{1.5cm} 
\begin{center}
\includegraphics[clip=true,width=0.8\columnwidth]{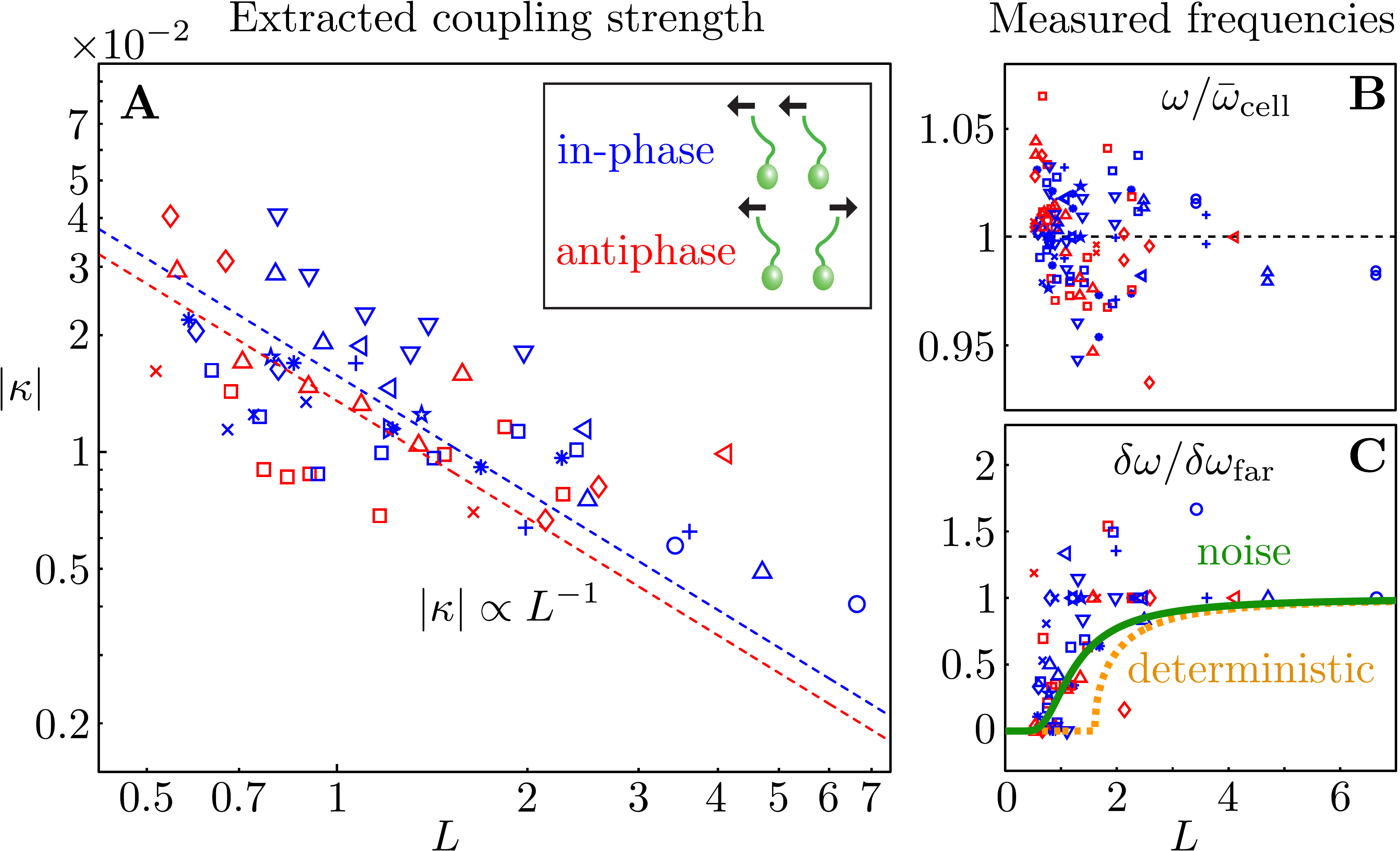}
\end{center}
\caption{{\bf Coupling strength.} {\bf (A)} Dimensionless interflagellar coupling strength $\kappa=\epsilon/\bar{\omega}$ as a function of the scaled 
spacing $L = d/l$ (log-log scale). The dotted lines represent fits of the form $|\kappa|=k\times L^{-1}$ with $k=0.016$ (in-phase) 
and $k=0.014$ (antiphase). {\bf (B)} Measured beat frequency $\omega / \bar{\omega}_{\text{cell}}$ of each flagellum, nondimensionalised by the average value for that cell across several videos. {\bf (C)} Measured frequency difference $\delta \omega / \delta \omega_{\text{far}}$ as a function of spacing $L$. The curves represent the predictions based on the average extracted model parameters in the absence (orange) and presence of noise (green). Symbols represent different pairs of cells, with the in-phase (blue) and antiphase (red) configurations shown.}
\end{figure*}

\newpage
\begin{figure*}
%\hspace{3cm} 
\begin{center}
\includegraphics[clip=true,width=\columnwidth]{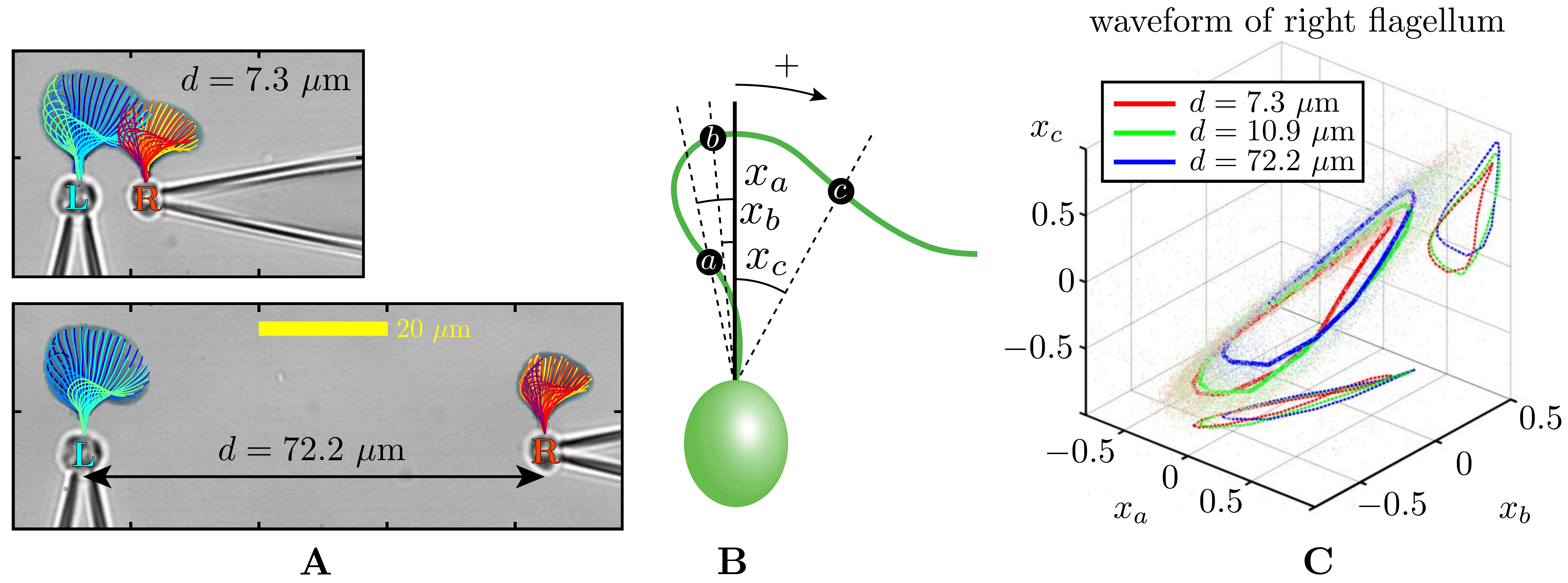}
\end{center}
\caption{\textbf{Waveform characteristics.} {\bf (A)} Logarithmically-scaled residence time plots of the entire flagella. The displayed waveforms correspond to 1~ms time intervals over several successive flagellar beats. {\bf (B)} Angles $x_a$, $x_b$, $x_c$ (in radians) measured and {\bf (C)} their characteristic 3D trajectories. Results are shown for the right flagellum, corresponding to three different interflagellar spacings. As the spacing $d$ is increased, the flagellar waveform exhibits a systematic change.}
\end{figure*}

\clearpage
%%%%%%%%%%%%%%%%%%%%%%
%%%%%%%%%%%%%%%%%%%%%%
%%%%%%%%%%%%%%%%%%%%%%

\newpage
\section*{Materials and Methods}
\setcounter{figure}{0}
\renewcommand{\thefigure}{S\arabic{figure}}
\setcounter{equation}{0}
\renewcommand{\theequation}{S\arabic{equation}}

\subsection*{Cell growth and imaging.}
\textit{Volvox carteri }  f. \textit{nagariensis} (strain EVE) were grown axenically in Standard \textit{Volvox} Medium (SVM)\cite{Kirk:1983uq} with sterile air bubbling, in a growth chamber (Binder, Germany) set to a cycle of 16~h light (100 $\mu$Em$^{-2}$s$^{-1}$, Fluora, OSRAM) at $28^{\circ}$C and 8~h dark at $26^{\circ}$C. Individual biflagellate cells were extracted from \textit{Volvox} colonies using a cell homogeniser, isolated by centrifugation with Percoll (Fisher, UK), and inserted into a $ 25 \times 25 \times 5$~mm glass observation chamber filled with fresh SVM. Cells were captured using micropipettes and oriented so that their flagellar beating planes coincided with the focal plane of a Nikon TE2000-U inverted microscope. Motorised micromanipulators (Patchstar, Scientifica, UK) and custom-made stages facilitated accurate rotation and translation of the cells. The flow field characterisation and pairwise synchronization analyses were imaged using a $40 \times$ Plan Fluor objective lens (NA 0.6). A higher magnification $63 \times$ Zeiss W Plan-Apochromat objective lens (NA 1.0) was used to conduct separate experiments for the waveform analysis. For each experiment, we recorded movies with a high-speed video camera (Fastcam SA3, Photron, USA) at 1000~fps under bright field illumination.

{\it One cell.}---Spatiotemporal analysis of the flow field associated with one cell was achieved through seeding the fluid with $0.5~\mu$m polystyrene microspheres (Invitrogen, USA) at a volume fraction of $2 \times 10^{-4}$. We recorded 104~s long movies, each one corresponding to approximately 3500 flagellar beats. The time-dependent velocity field was reconstructed using an open source particle image velocimetry\cite{Raffel:PIV} toolbox for MATLAB (MatPIV). 

{\it Two cells.}---For each pair of \textit{Volvox} cells, we investigated the synchronization properties as a function of interflagellar spacing. A number of movies were taken at various separations (non-monotonically varied). In many cases we also rotated pipette P$_1$ (see Fig.~1B) so that the flagella were beating in the same plane but opposite directions. There are two such ``antiphase'' configurations possible, in which the flagella beat \textit{towards} and \textit{away} from one another respectively. Both of these states are referred to as \textit{antiphase} in the extraction of parameters in Fig.~3 and Fig.~S2.

\newpage
\subsection*{Phase extraction.}
Movies of hydrodynamically interacting flagella were first processed by subtracting a 30 frame running average. Median filtering was undertaken using $3\times3$ pixels$^2$ regions. We defined small interrogation regions for each cell (red and blue quadrilaterals in Figs.~S1A--C), so that the respective flagella passed through precisely once per beat. Recording the passage times between beats (see Figs.~S1D--F) allowed reconstruction of the flagellar phase $\phi_{1,2}\in[0,2\pi]$. The time-dependent interflagellar phase difference $\Delta(t) = (\phi_1 - \phi_2)/2\pi$ was used to characterise the synchronization properties of the two cells.

\begin{figure}[h!]
%\hspace{-1cm}
%\includegraphics[width=16.5cm]{figures/signal.pdf}
\begin{center}
\includegraphics[width=\columnwidth]{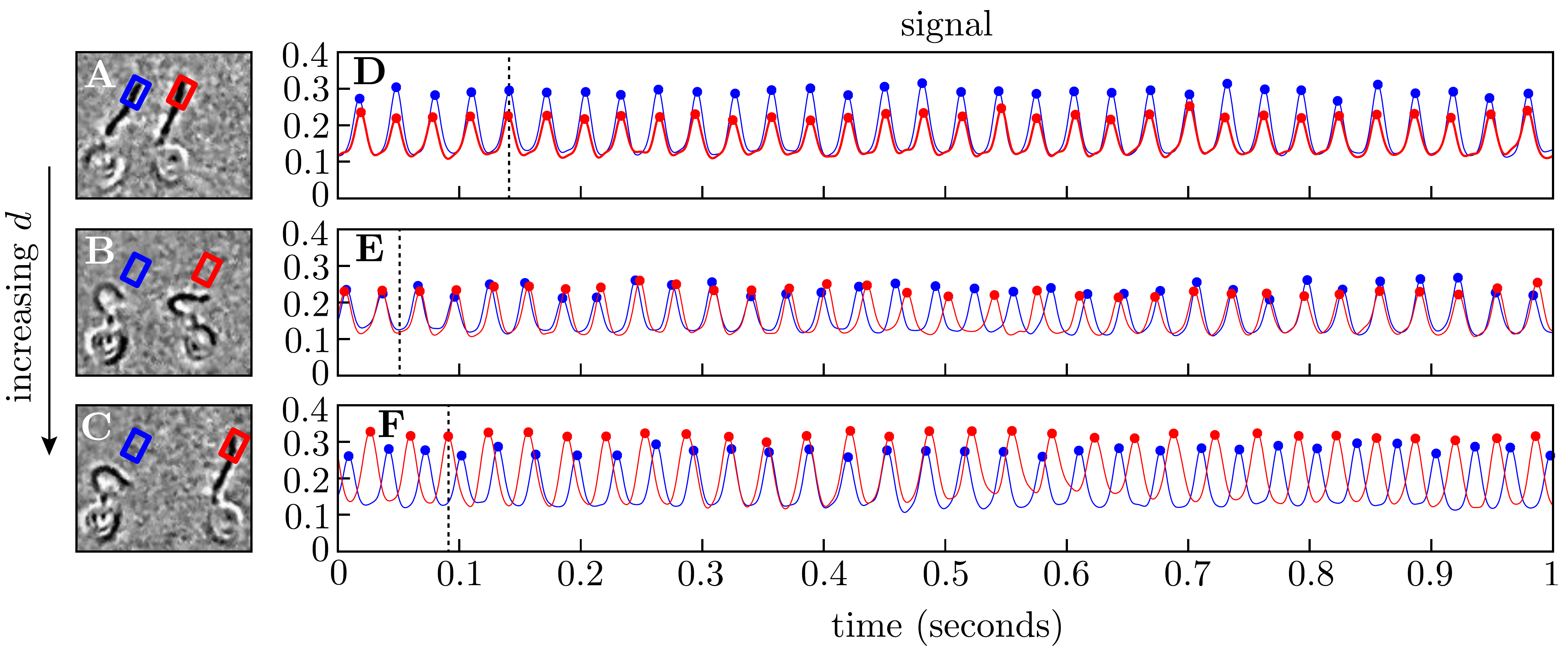}
\end{center}
\caption{\textbf{Phase extraction}. {\bf (A--C)} Snapshots from three different processed movies showing the same cells at different interflagellar spacings ($d=14.2$, $20.4$, $28.2~\mu \text{m}$).  {\bf (D--F)} The signal extracted from the interrogation regions is used to reconstruct the flagellar phases. For each row, the frame on the left corresponds to the time indicated by the dashed line.}
\label{fig:signal}
\end{figure}

\subsection*{Additional model parameters.}
The stochastic Adler equation was used to model the dynamics of $\Delta(t)$ as described in \cite{Goldstein:2011_emergence}.  
Figures~S2A--B show the amplitude $C_0$ of the autocorrelation function of $\Delta$ and the values of the average synchronous period $\tau_{\text{sync}}$. Fluctuations of the phase-difference $\Delta$ about the synchronized states are well described by Gaussian distributions, with variances $C_0$ proportional to the interflagellar spacing $L$.  The coupling strength $\epsilon$ exhibited excellent agreement with the hydrodynamic predictions. Figures~S2C--D show the dependence of the effective temperature $T_{\text{eff}}/\bar{\omega}$ and intrinsic frequency difference $\delta\nu/\bar{\omega}$ as a function of $L=d/l$ for every pair of measured cells. 

\begin{figure}[h!]
%\hspace{0cm}
%\includegraphics[width=14cm]{figures/model_parameters.pdf}
\begin{center}
\includegraphics[width=\columnwidth]{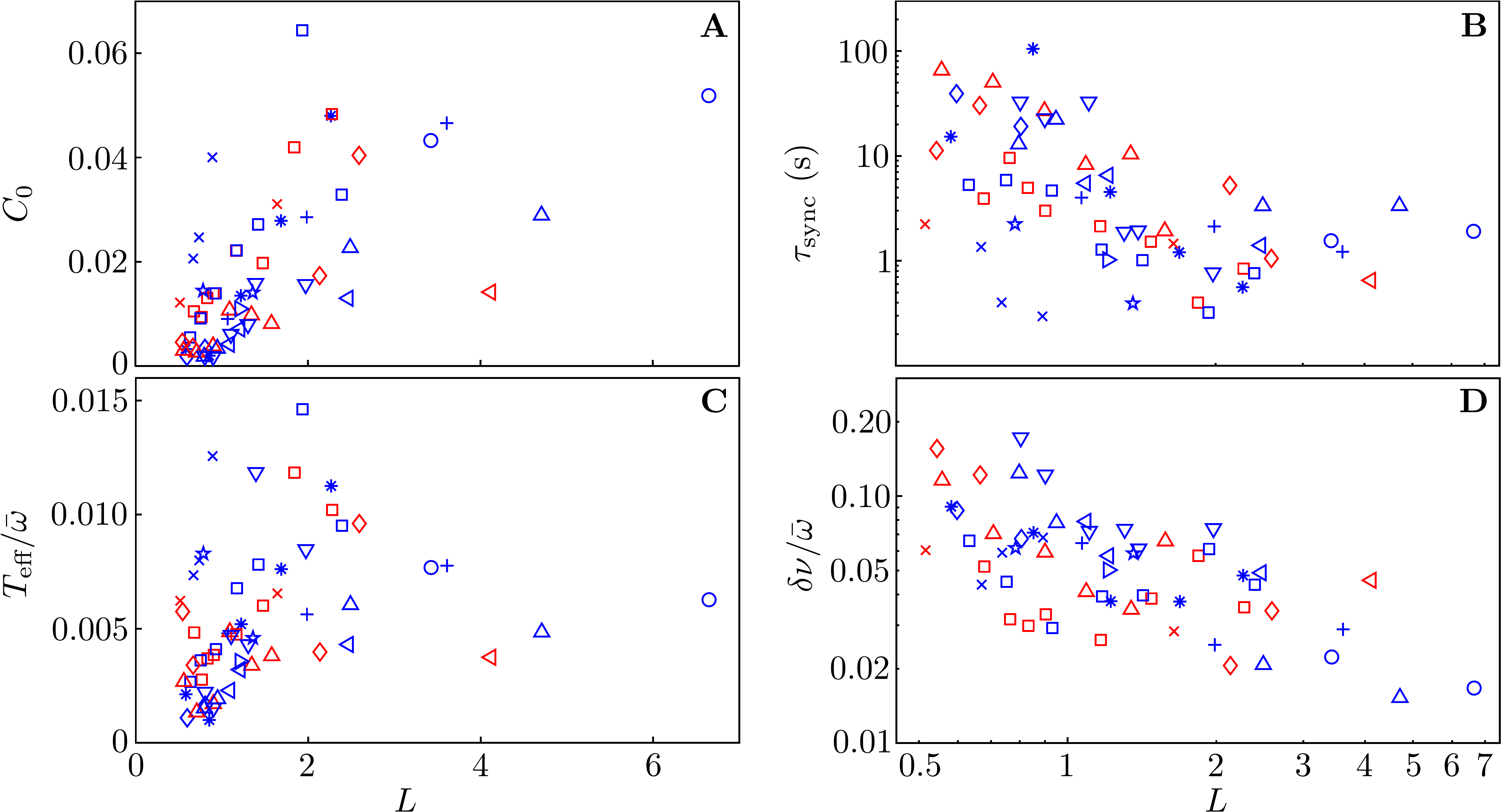}
\end{center}
\caption{\textbf{Model parameters}. Two of the experimental observables {\bf (A)} $C_0$ and {\bf (B)} $\tau_{\text{sync}}$, and the two additional model parameters {\bf (C)} $T_{\text{eff}}/\bar{\omega}$ and {\bf (D)} $\delta \nu / \bar{\omega}$ are shown as a function of interflagellar spacing for all experiments conducted.}
\label{fig:model_param}
\end{figure}

\subsection*{Waveform analysis.} 
High-speed video recordings were image-processed and enhanced in MATLAB, and flagellar waveforms extracted as described in \cite{wan14}. 
Figure~4A shows the waveforms over several successive beats, measured at 1~ms time intervals. 
Logarithmically-scaled residence time plots for the entire flagella are also shown. Each flagellum was divided into four segments of equal arc length, with the three interior points (moving from proximal to distal end) making angles $x_a$, $x_b$ and $x_c$ (measured in radians) respectively with the cell body axis (see Fig.~4B). Three-dimensional scatter plots displaying the extracted coordinates $(x_a,x_b,x_c)$ from every frame of the videos are shown in Fig.~4C. Bringing the cells closer together results in a characteristic shift in the waveform.
\begin{figure}[h!]
\begin{center}
\includegraphics[width=\columnwidth]{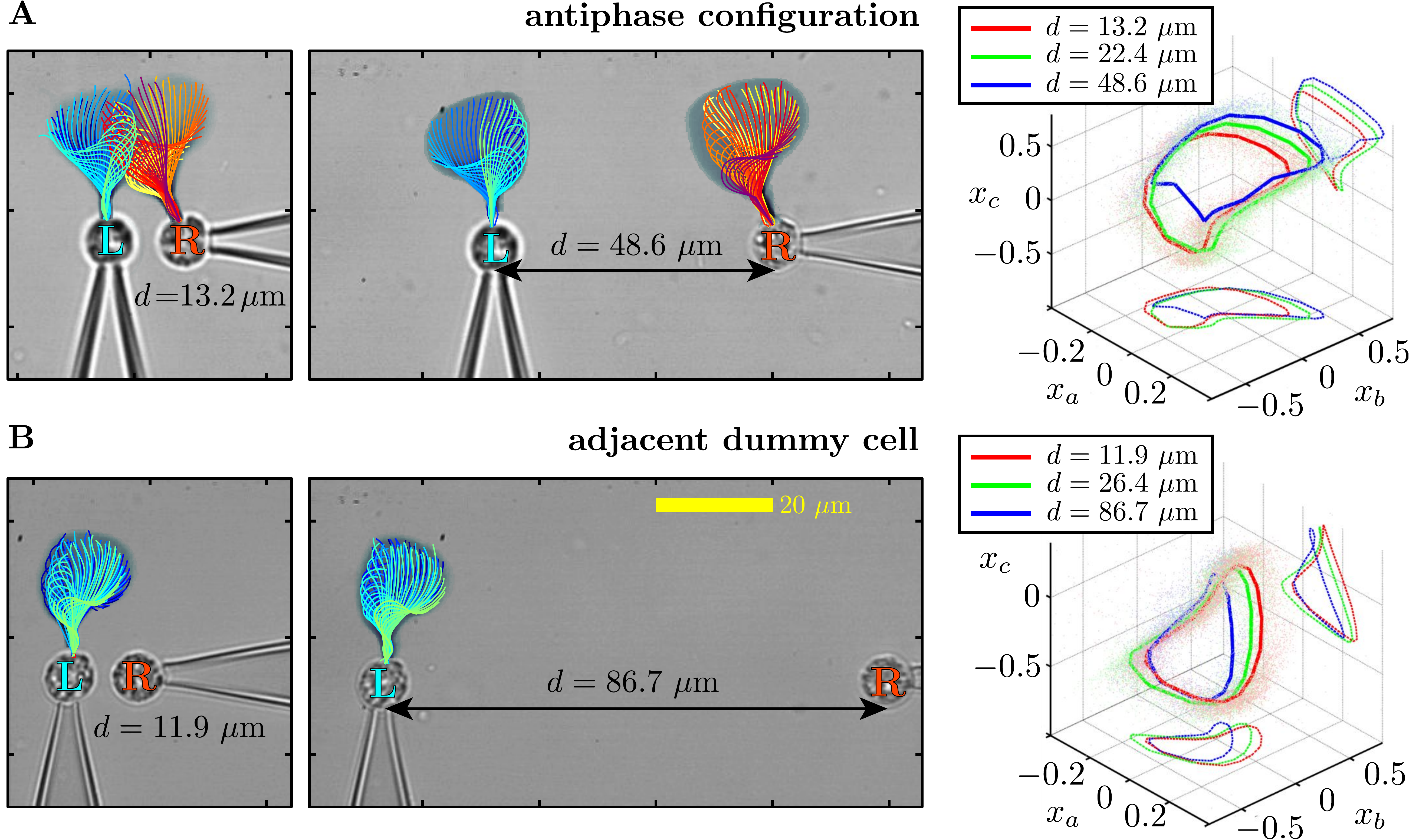}
\end{center}
\caption{ {\bf Waveform analysis}. Flagellar filaments are tracked for cells in the {\bf (A)} antiphase state, as well as {\bf (B)} the situation in which one of the cells does not possess a flagellum (dummy cell). For each configuration, the waveform of the left cell is analysed at three different cell--cell separations.}
\label{waveform_supplementary}
\end{figure}

%For the antiphase: 13.2, 22.44, 48.62 um
%For the dummy: 11.88, 26.4, 86.68 um

\subsection*{Resistive force theory.} 
Through digital waveform tracking we are able to precisely resolve the spatiotemporal dynamics of a beating flagellum.
Figure~S4A shows snapshots of a typical flagellum captured over a full beat cycle, superimposed at $2\,$ms intervals, together with measured instantaneous velocities along the filament.
With resistive force theory (RFT), the results of the tracking procedure are used to derive estimates for the forces produced by the flagellum. First proposed by Gray and Hancock \cite{GrayHancock}, RFT considers the anisotropic drag experienced by a long rod-like flagellum moving through a viscous fluid, and assumes that each unit segment of the flagellum experiences a local drag that is proportional to its local instantaneous velocity. 
The force density $\bm{f}$ along the flagellum is approximated by
\begin{equation}
\bm{f} = C_{\perp} \bm{u}_{\perp} + C_{||} \bm{u}_{||}, 
\end{equation} 
where the constants of proportionality, $C_{\perp}$ and $C_{||}$, are the normal and tangential resistance coefficients respectively, and are readily computable from experimental data.
We chose $C_{\perp}$ and $C_{||}$ according to \cite{Lighthill}, with $C_{\perp}=4\pi\mu/(\ln(0.18\lambda/a)+0.5)$ and $C_{||}=2\pi\mu/(\ln(0.18\lambda/a)-0.5)$, with an aspect ratio $\lambda/a=80$.

The total instantaneous force $\bm{F}(t)$ produced by the flagellum is given by $\int_0^l\bm{f}(s,t)\,ds$, where $l$ is the total length of the flagellum and $s$ its arclength parameterisation.
In Fig.~S4B we plot the normal (blue) and tangential (red) components of $\bm{f}$, for characteristic power and recovery stroke waveforms (solid \& dotted lines respectively). 
To construct a limit-cycle representation of the cyclic force variation, we define an effective centre-of-mass for the flagellum, $\bm{x}(t)=\sum_i^N(|\bm{f}_i|\bm{x}_i(t))/\sum_i^N(|\bm{f}_i|)$, averaging over all $N$ discretized force vectors $\bm{f}_i$ applied at points $\bm{x}_i$ along the flagellum. 
Figure~S4C depicts the trajectories of integrated force $\bm{F}$ in this coordinate representation (red arrows).
An average limit cycle representation (black arrows) is obtained from measurements taken from $30$ beats: resultant force directions are seen to vary continuously along the cycle. 
The RFT result is seen to overestimate the force production during the recovery stroke, where the assumption of locality breaks down.

\vspace{1cm}
\begin{figure}[h!]
\begin{center}
\includegraphics[width=\textwidth]{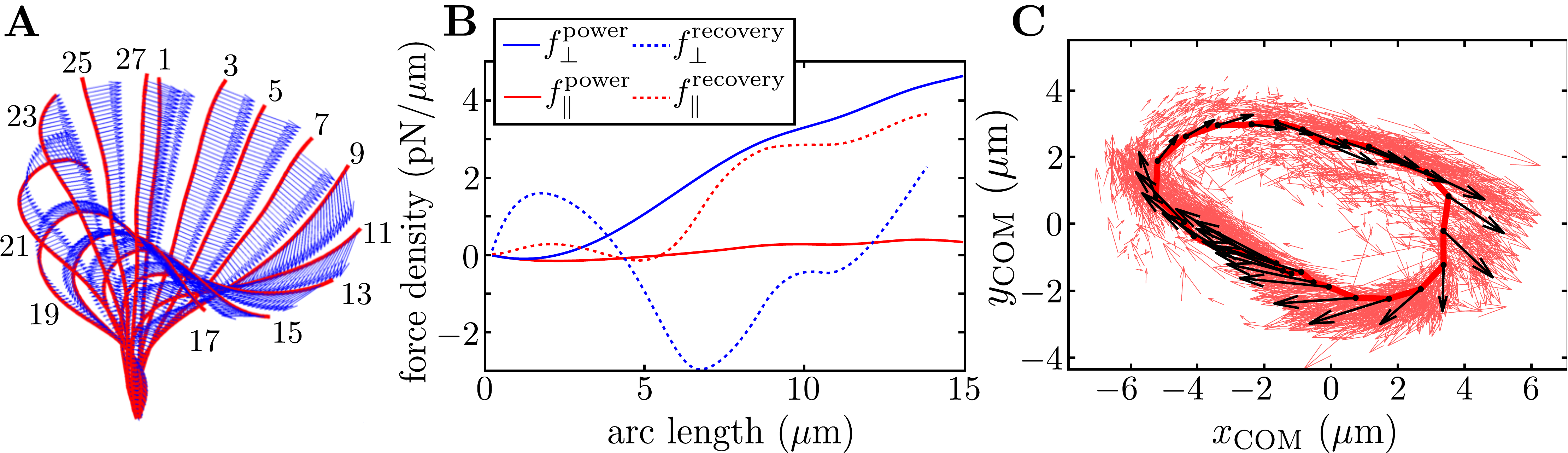}
\end{center}
\caption{\textbf{Resistive force theory analysis}. {\bf (A)} Instantaneous velocity distribution along the flagellum during one complete beat cycle (indexed by frame number, imaged at 1000~fps). {\bf (B)} Components of integrated force produced by a flagellum executing characteristic power and recovery strokes. {\bf (C)} Integrated vector forces $\bm{F}(t)$ shown localised at centre-of-mass coordinates $\bm{x}(t)$ (red), evolve cyclically around an average trajectory. Scale bar: $|\bm{F}|/8\pi\mu\sim1910~\mu \text{m}^2/\text{s}$.}
\label{RFT}	
\end{figure}

\newpage
\subsection*{Flagellar flow fields.}

We present here the time-dependent flow fields associated with one individual beating flagellum, from which the average in Fig.~2B is constructed. Figure~S5 illustrates the fluid velocity direction (vector field) and magnitude (colour) at various stages through one flagellar beat. The velocity magnitude during the flagellar power stroke (frames 1,2,6--8) is larger than during the recovery stroke (frames 3--5), giving rise to a non-zero net flow. In order to characterise these experimental results, we fit the flow field in each frame using a point force (Stokeslet) of unknown position $\bm{x}_0$, magnitude and direction $\bm{F}$. The corresponding velocity is given by
\begin{equation}
u_i = \frac{F_j}{8 \pi \mu} \big( \delta_{ij}/r + r_i r_j/r^3 \big),
\end{equation}
where the vector $\bm{r} = \bm{x} - \bm{x}_0$, $r = |\bm{r}|$, $i\in\{x,y\}$, and $\delta_{ij}$ is the Kronecker delta. The results of this fitting procedure for individual frames are illustrated in Fig.~S5. The average trajectory $\bar{\bm{x}}_0(t)$ executed by the Stokeslet over approximately 1000 beats is represented by the closed white curve. 

\begin{figure}
\begin{center}
\vspace{-1cm}
\includegraphics[width=\textwidth]{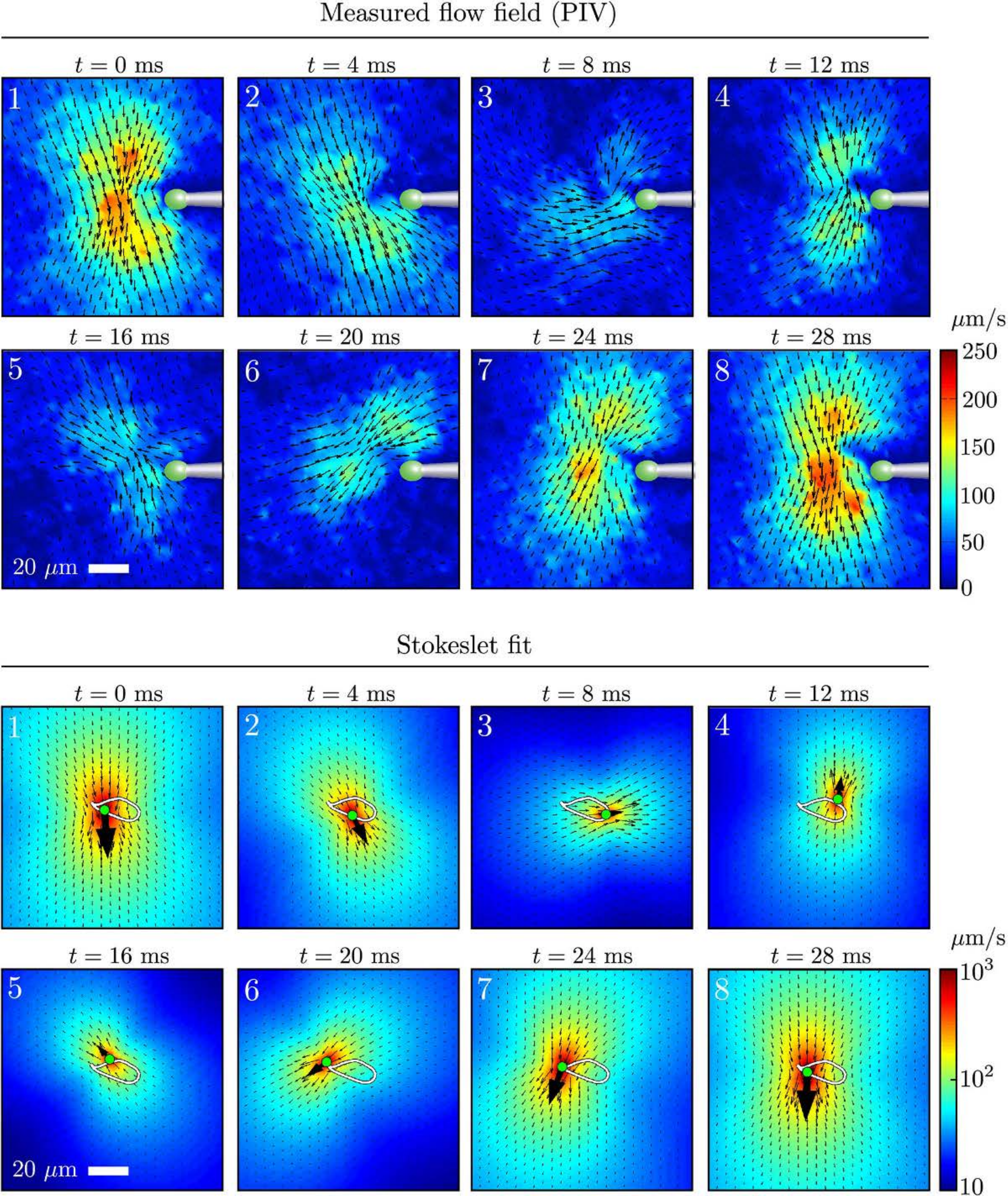}
\end{center}
\caption{\textbf{Time-dependent flow fields}. Instantaneous fluid velocity corresponding to various stages during one representative flagellar beat. Also shown are the fitted flow fields for each frame, corresponding to the application of a point force on the fluid. Very good qualitative agreement can be seen.}
\label{beating_cycle}
\end{figure}

\begin{figure}
%\hspace{1cm} 
\begin{center}
\includegraphics[width=0.9\textwidth]{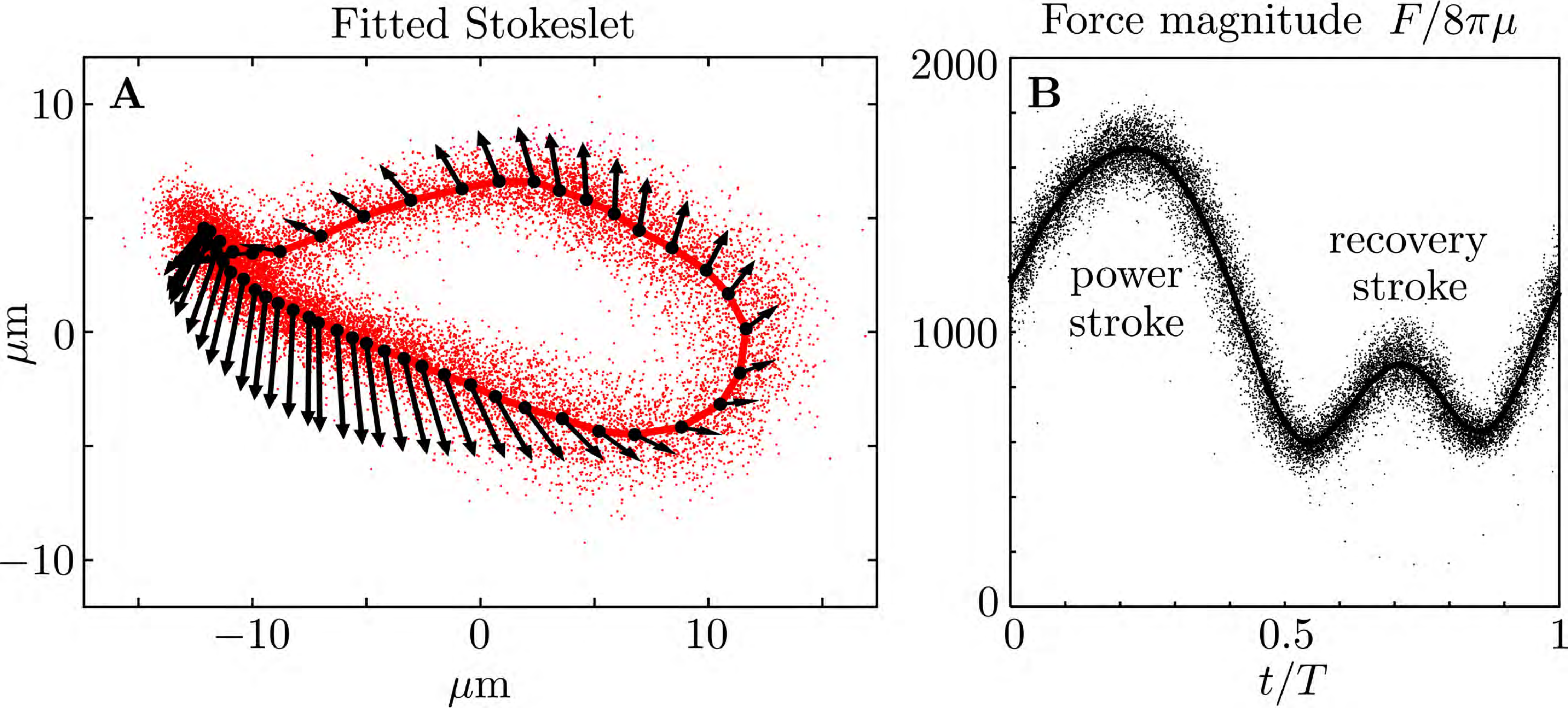}
\end{center}
\caption{\textbf{Force amplitude of flagellum}. {\bf(A)} The fitted Stokeslet is shown at evenly-spaced times throughout the average flagellar beating cycle. The red dots indicate the Stokeslet position extracted from every frame. {\bf(B)} Amplitude of the fitted point force as a function of time throughout the flagellar beat.}
\label{beating_cycle_Stokeslet}
\end{figure}
Figure~S6A shows $\bar{\bm{x}}_0(t)$ (solid red line) together with a scatter plot of $\bm{x}_0(t)$ from individual frames (red dots). The black arrows along the average cycle illustrate the average position, orientation and relative magnitude of the Stokeslet at evenly-spaced times along the \textit{average cycle}. Importantly, the orientation of the point force does not coincide with its direction of motion, a fact to be expected given the anisotropic drag on the flagellum. Fig.~S6B shows the magnitude of the fitted Stokeslet over all beats. The amplitude of this force exhibits very strong periodic variations, and is well approximated by $F(t)/8\pi\mu = A_0\left(1+A_1\sin(2\pi\,t/T)\right)$ with $A_0\simeq1076\,\mu$m$^2/$s and $A_1\simeq0.56$. It is encouraging to see that the amplitude of this force is similar to the value calculated using resistive force theory earlier, though it should be noted that these results correspond to two different cells.

From the distribution of beating periods, we are able to directly estimate the effective internal noise to be $\langle T_{\text{eff}}/\bar{\omega} \rangle= 0.002$, a value consistent with the extracted model parameter $\langle T_{\text{eff}}/\bar{\omega} \rangle = 0.005 \pm 0.003$.

\subsection*{Proximity to pipettes.}

In order to study the dynamics of hydrodynamically coupled flagella, the two cells were held using orthogonal glass pipettes. This geometry allowed us to investigate both in-phase and antiphase configurations for the same pair of cells, through the simple rotation of one pipette. At the same time, however, this meant that the two cells were held from different directions with respect to their flagella, and that one of the two pipettes was oriented along the direction of the flagellar power stroke, which is the main flow direction. This can cause two problems. Firstly, the flow field of a cell held by the side could be significantly different from that presented in Fig.~2. Secondly, the holding pipettes could distort the scaling of the flagellar flow with cell-to-cell separation from the $\sim1/r$ scaling presented in Fig.~2C.
We investigated these problems with the series of experiments in Fig.~S7. One cell was held by a pipette opposite to the flagellar apparatus (Fig.~S7A) and the flow field measured. A second micropipette was then moved progressively closer, eventually to the point of contact with the cell (Fig.~S7D). It is clear that the second pipette affects the flow, but mostly in the region between the two pipettes. 

\begin{figure}[h!]
\begin{center}
\includegraphics[width=\textwidth]{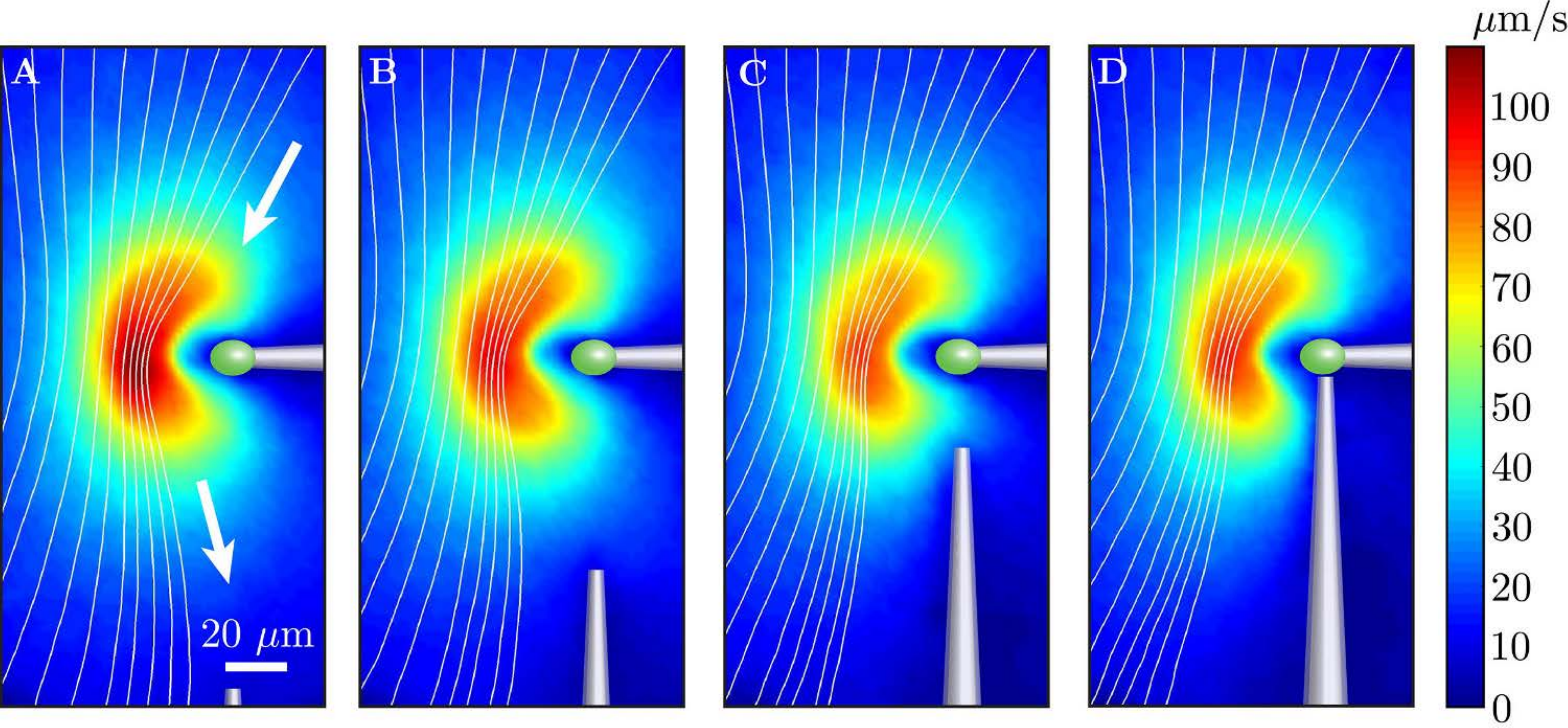}
\end{center}
\caption{\textbf{Effect of nearby pipette}. The time-averaged flow field associated with one captured cell is measured as a second pipette slowly approaches. This demonstrates that the precise angle from which the cell is held by the micropipette has very little effect on the resultant flow field.}
\label{flow_field_pipette}
\end{figure}

Let us consider the region upstream of the cell (above the cell in Fig.~S7D). For a cell held from the side, this is the region where the other cell will be. Here the flow is only minimally affected, with an average relative change between Figs.~S7D and A  below $8\%$. A large contribution is represented simply by a $\sim7\%$ decrease in flow speed. Taking this decrease into account, the average relative change is about $5\%$. As a result, these experiments allow us to consider the flow generated by a cell held from the side as identical to that generated by a cell held from the back, at least in the region of interest to our experiments.
By comparing the flows for different positions of the second pipette, we can also quantify its effect on the flow field that would be experienced by the second cell. For each pipette configuration, this can be estimated as the relative difference between the unperturbed and the perturbed flows in the region where the flagella of the second cell would be, here considered to be a $10\times10\,\mu$m region $20\,\mu$m to the left of the tip of the incoming pipette. The difference ranges from $\sim5\%$ to $\sim10\%$ and $\sim13\%$ for Figs.~S7B,C,D respectively (in the last case we choose a position approximately $10\,\mu$m below and $20\,\mu$m to the left of the pipette tip). These represent the typical error contributions from neglecting, as we have done in the text, the influence of the pipettes on the flows generated by the cells.

\subsection*{Minimal model with variable forcing.}
We used the Stokeslet approximation to the flow field of an isolated cell in Fig.~S6, to test the effect of force modulation on synchronization within the class of minimal models which abstract the beating flagellum as a sphere driven along a closed orbit \cite{Niedermayer:2008fk,Uchida:2011kx,Uchida:2012fk}.
We simulated two spheres of radius $a=0.75\,\mu$m in an unbounded fluid of viscosity $\mu=10^{-3}\,$Pa\,s, driven along coplanar circular orbits of radius $r_0=8\,\mu$m by a force $F(\phi)/8\pi\mu = A_0(1+A_1\sin(\nu\phi+\phi_0))$ tangential to the orbit, with $A_0=1076\,\mu$m$^2/$s and $A_1=0.56$. Notice that this corresponds to assuming that the point forces in Fig.~S6A are tangent to the cycle. The value $a=0.75\,\mu$m was chosen to ensure that the orbital frequency matched the mean value observed experimentally. The orbits were separated by $d=20\,\mu$m and had a radial stiffness with spring constant $\lambda$. The limit $\lambda\to\infty$ corresponds to rigid prescribed trajectories (holonomic constraint).  For each value of $\lambda\in\{1\,\textrm{pN}/\mu\textrm{m},5\,\textrm{pN}/\mu\textrm{m},\infty\}$ we ran 5 sets of simulations, corresponding to $\nu\in\{0,1,2\}$ and $\phi_0\in\{0,\pi/2\}$. Choosing $\nu=2$ is equivalent to modulating the driving force with the experimental amplitude but at a frequency double the experimental one. Although this is not what we observed, it is still interesting to consider, since in this configuration it is the frequency that most contributes to synchronization through force modulation. 

\begin{figure}[h!]
\begin{center}
\includegraphics[width=\textwidth]{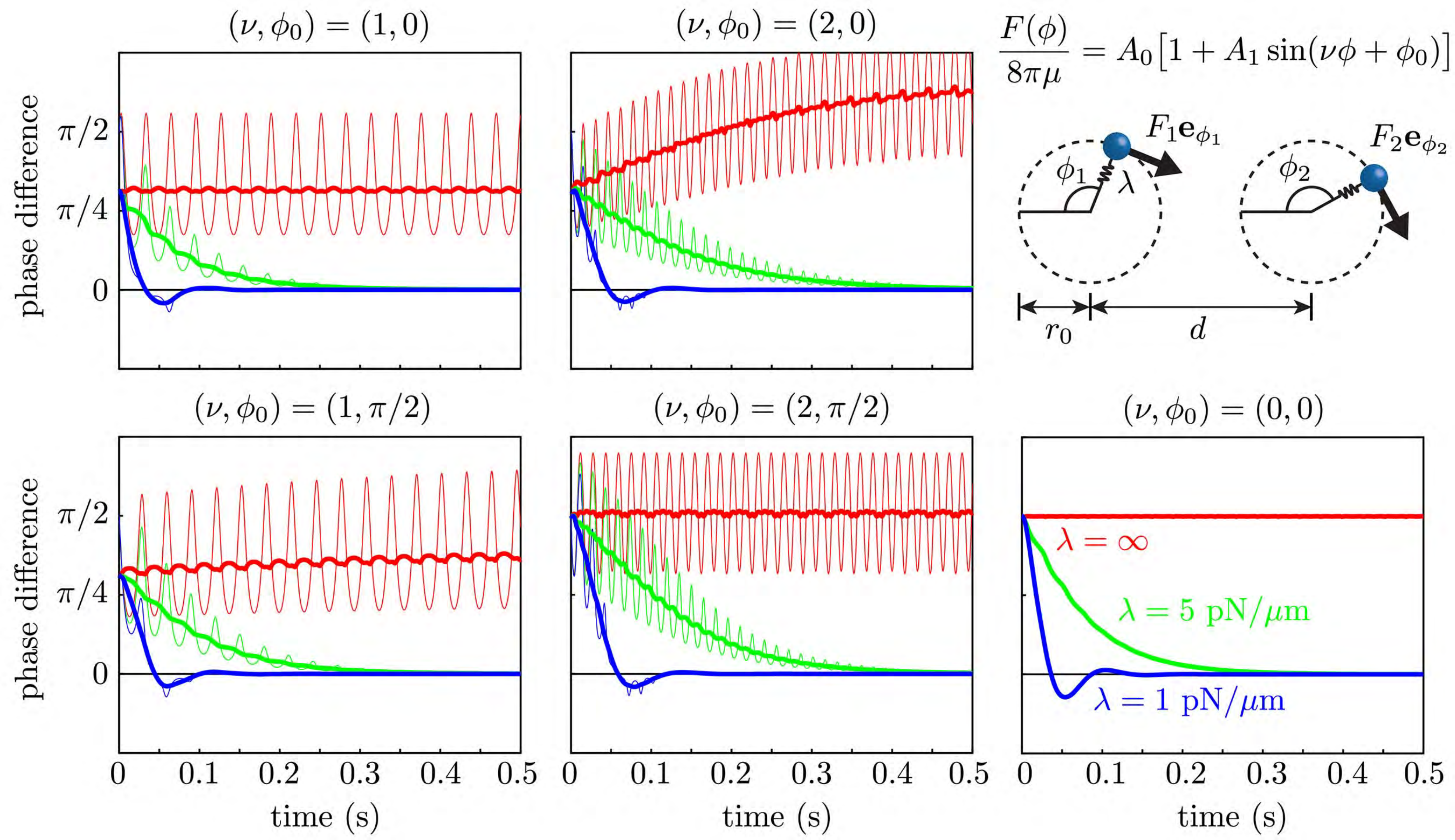}
\end{center}
\caption{\textbf{Effect of force modulation}. Evolution of the phase difference $\delta = \phi_1 - \phi_2$ among two identical model oscillators, each composed of a sphere driven around a circular trajectory by a tangential driving force. The trajectories each possess a radial stiffness $\lambda$. Smaller values of $\lambda$ yield rapid convergence towards synchrony ($\delta = 0$), in a manner essentially independent of the functional form of the driving force. Parameters used are given by $a=0.75\,\mu$m, $r_0=8\,\mu$m, $d=20\,\mu$m, $A_0=1076\,\mu$m$^2/$s and $A_1=0.56$.}
\label{simulations_1}
\end{figure}

As a consequence of the phase-dependent driving force, the geometric phase $\phi_i$ of an {\it individual isolated} oscillator does not evolve at a constant rate in time. We thus chose to rescale the phase $\Phi = \Phi(\phi)$ so that in the absence of hydrodynamic interactions, $\dot{\Phi} = 2\pi/T = \,$constant. Both the {\it geometric} phase difference $\delta = \phi_1 - \phi_2$ (thin curves) and its {\it rescaled} value $\delta_{\text{rescaled}} = \Phi_1 - \Phi_2$ (thick curves) is shown for each simulation in Fig.~S8. These results show clearly that {\it within the boundary} of the model we are considering, the two oscillators synchronize through a coupling between hydrodynamic stresses and orbit compliance \cite{Niedermayer:2008fk} with no noticeable effect from force modulation. \\

Repeating the simulations with a stiffness derived from the flagellar bending rigidity as in the main text, $\lambda=0.05\,\textrm{pN}/\mu\textrm{m}$, radius $a=0.1\,\mu$m, and  reducing the force amplitude to $A_0=143\,\mu$m$^2/$s to keep the revolution frequency at the experimental value, yields the results in Fig.~S9. Again, the synchronization is achieved only through interaction between hydrodynamic stresses and orbit compliance. 

\begin{figure}[h!]
\begin{center}
\includegraphics[width=\textwidth]{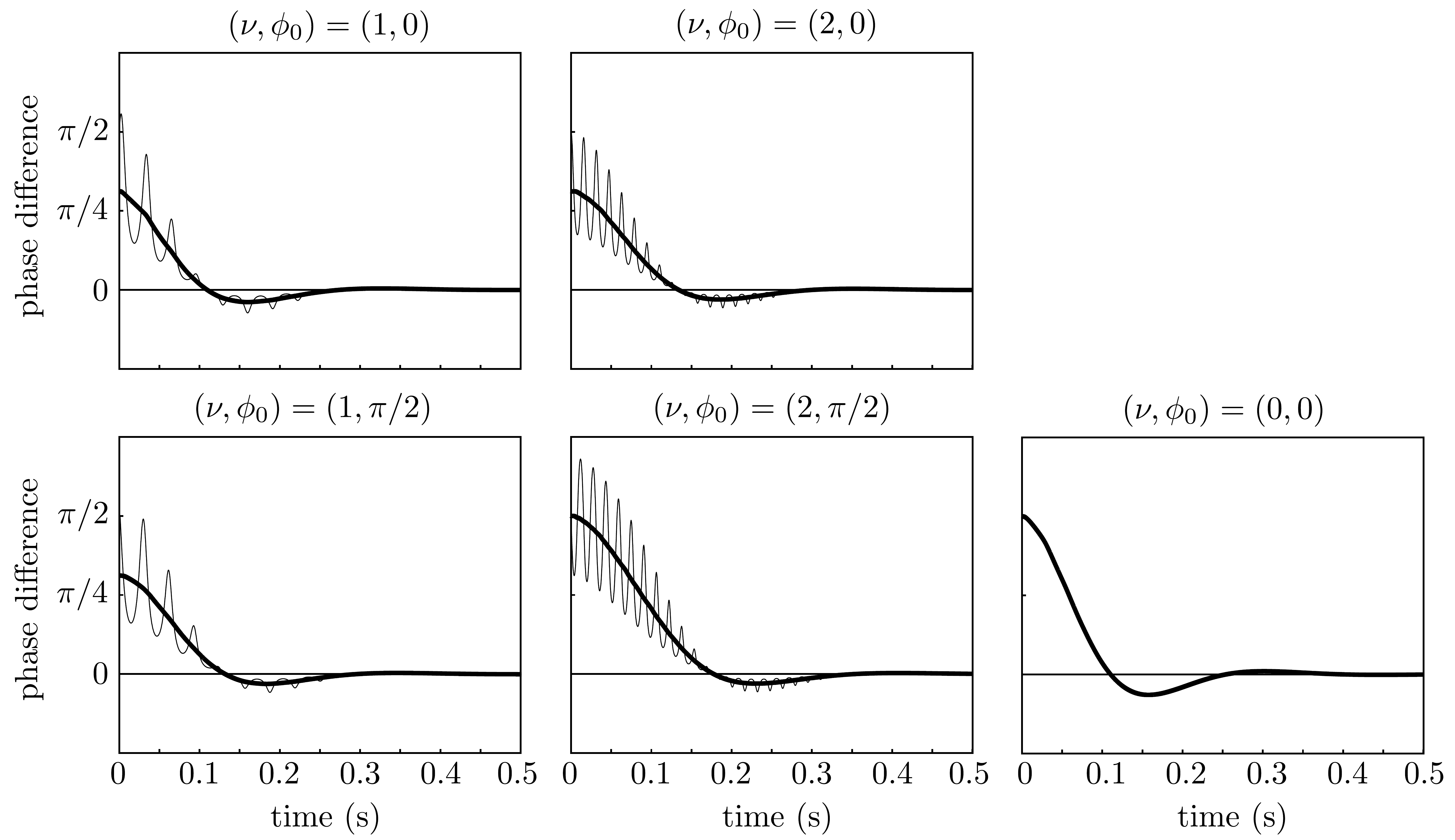}
\end{center}
\caption{\textbf{Effect of force modulation}. Re-run of the simulations in Fig.~S8 with properties inspired by real flagella.}
\label{simulations_2}
\end{figure}

\end{document}